\documentclass[twocolumn,showpacs,preprintnumbers,amsmath,amssymb]{revtex4}

\usepackage{graphicx}
\usepackage{dcolumn}
\usepackage{bm}

\begin{document}
\widetext

\title{Comment on "Precision global measurements of London penetration depth in FeTe$_{0.58}$Se$_{0.42}$"}

\author{T. Klein$^{1}$,  P. Rodi\`ere$^1$, and C.Marcenat$^{2}$}

\address{$^{1}$  Institut N\'eel, CNRS and Universit\'e Josepph Fourier  BP166, F-38042 
Grenoble, France}
\address{$^2$ SPSMS, UMR-E9001, CEA-INAC/ UJF-Grenoble 1, 17 rue des 
martyrs, 38054 Grenoble, France}

\begin{abstract}
Cho {\it et al.} [Phys. Rev. B, {\bf 84}, 174502 (2011)] have reported on the temperature dependence of the London penetration depth deduced from Tunnel Diode Oscillator (TDO) measurements in optimally doped Fe(Se,Te) single crystals.  According to their analysis, these measurements  chould suggest a nodeless two-gap pairing symmetry with strong pair breaking effects. However, to reach this conclusion, the authors fit the temperature dependence of the superfluid density with a two band {\it clean} limit model which is incompatible with the presence of strong pair breaking effects, deduced from the $T^n$ temperature dependence of the London penetration depth below $T_c/3$. Moreover they claim that their results are also ruling out the suggestion that surface conditions can significantly affect the TDO data but this conclusion is based on one very specific damaging process, and is  completely ignoring the large dispersion in the previously published TDO data.  \end{abstract}

\pacs{74.60.Ec, 74.60.Ge}  
\maketitle

In a recent article Cho {\it et al.} \cite{Cho} reported on the temperature dependence of the London penetration depth deduced from Tunnel Diode Oscillator (TDO) measurements in a series Fe(Se,Te) crystals close to optimal doping. TDO is a powerful technique to determine very accurately the temperature dependence of the variation of the penetration depth ($\Delta\lambda(T)$), especially for $T \rightarrow 0$, and hence to obtain valuable informations on  the low energy excitations present in the system. All the measurements performed by several groups on Fe(Se,Te) samples from different origins agree for instance on the fact that, at low temperature, $\Delta\lambda(T)$ can be well described by a power law $\Delta\lambda(T) \propto T^n$  with $n \sim 2.0 - 2.3$  \cite{Cho,Kim,Sarafin,Klein}. This power law suggests  the presence of  pair breaking effects, as expected in dirty d-wave superconductors or in the case of interband scattering in s$\pm$ superconductors \cite{Gordon}. 

However, to obtain a complete description of the gap structure it is necessary to determine the temperature dependence of the normalized superfluid density ($\rho_s$) on the entire temperature range. As $\rho_s(T)=1/(1+\Delta\lambda(T)/\lambda_0)^2$, both the amplitude of $\Delta\lambda(T)$ and the zero temperature penetration depth $\lambda_0$ have to be determined precisely to obtain reliable $\rho_s(T)$ data. Nevertheless, very different $\Delta\lambda$ values have been reported in samples with very similar $T_c$ values (optimally doped samples) by the different groups involved in the study of Fe(Se,Te). Indeed,  taking for instance $T=5$ K for comparison purpose, the different $\Delta\lambda(5K)$ values are varying from  $\sim 35$ nm for the Bristol group \cite{Sarafin} to $\sim 130$ nm for the Grenoble Group \cite{Klein} and values ranging from $\sim 30-50$ nm in \cite{Cho} to $\sim 110$ nm in \cite{Kim} have been reported by the Ames group. It is hence of fundamental importance to understand the origin of this dispersion to obtain unambiguous $\rho_s(T)$ data.

A possible influence of edge roughness in iron based superconductors  has been pointed out by Hashimoto {\it et al.} \cite{Hashimoto} noting that $\Delta\lambda(T)$ can vary by a factor two from one sample to another in KFe$_2$As$_2$. In \cite{Cho}, the authors claim that they have ruled out the possibility that surface roughness can significantly affect the amplitude of $\Delta\lambda(T)$, and that they have shown that the temperature dependence of the superfluid density is consistent with a nodeless two gap pairing symmetry in the presence of strong pair breaking effects. However, we believe that the data have been over-interpreted  as (a) only one specific kind of disorder has been investigated, (b) they used the same $\lambda_0$ value to obtain $\rho_s$ in \cite{Cho} and \cite{Kim} even though the $\Delta\lambda(T)$ values differ by a factor $\sim 3$ and (c) the two gap {\it clean} limit model used to fit the $\rho_s(T)$ data is incompatible with the presence of strong pair breaking effects.

\paragraph{On the influence of edge roughness on $\Delta\lambda (T)$.} This possibility has been rejected in \cite{Cho} on the basis of measurements on sample "2-R" in which some roughness has been introduced by razor damaging and for which $\Delta\lambda (5K)$ (only) increases by $\sim 60 \%$ (and $A$ is subsequently only marginally modified when the  edges are cut back "as clean as possible"). Unfortunately, no structural information on the roughness introduced by the razor damaging is given but it is hard to believe that such a procedure could be characteristic of all kind of edge roughness. Indeed, the amplitude of the TDO signal will be sensitive to a roughness on the scale of $\lambda_0$ but would be only marginally affected by defects on much larger scales. Moreover, even if the dispersion in the $\Delta\lambda(T)$ values measured in \cite{Cho} remains reasonable ($\sim 30-50$ nm), the authors are completely ignoring the large dispersion in the values previously published (including their own data which differ by a factor $\sim 3$) stating that the values obtained in \cite{Cho} are "similar to other reports". The origin of this large dispersion remains an open question. The amplitude of $\Delta\lambda$ might be affected by various parameters such as sample inhomogeneities or disorder (see discussion below), microcracks, etc...., but the influence of edge roughness can {\it not} be excluded from the data presented in \cite{Cho}. 
\begin{figure}
\begin{center}
\resizebox{0.48\textwidth}{!}{\includegraphics{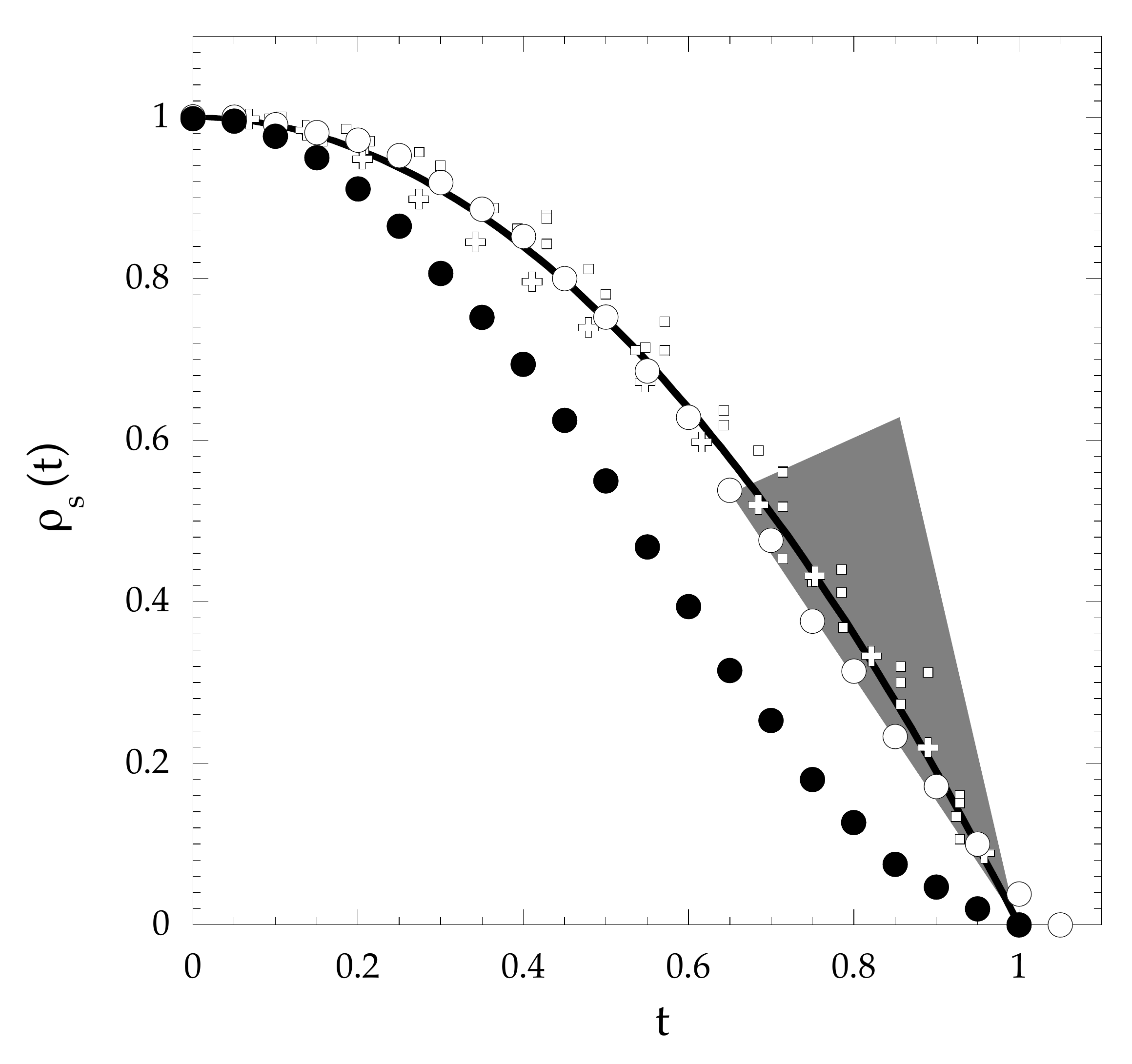}}
\caption{Temperature dependence of the superfluid density deduced from TDO in [2] (closed circles) and [1] (open circles) together with the one deduced from $H_{c1}$ measurements in [4] (open squares) and $\mu$SR data [9] (open crosses). The solid line is a $1-t^2$ dependence (with $t=T/T_c$) and the shaded area correspond to the $\rho_s$ values expected from the specific heat jump at $T_c$ (see text for details). }
\end{center}
\end{figure}

\paragraph{On the choice of the parameters.}  In a clean s-wave superconductor (that is for a mean free path $l$ larger than the coherence length $\xi$), $\lambda(0)$  depends only on the Fermi Surface properties \cite{Chandrasekhar}  and is hence independent of the sample quality. However, it is now well established that strong pair breaking effects are present in iron based superconductors and Fe(Se,Te) can not be discussed within the framework of those standard materials. Indeed, in contrast those later systems, all scattering events (and not only spin-flip scattering) are expected to be detrimental \cite{Onari} and the absolute value of the superfluid density is reduced even for $T \rightarrow 0$.  $\lambda(0)$ is then expected to depend strongly on $T_c$ (see for instance \cite{Gordon}) and,  concomitantly, $\Delta\lambda(5K)$ will be directly proportional to $\lambda(0)/T_c^2$ (for strongly reduced $T_c$ values \cite{Gordon}).  This close relationship between $\Delta\lambda$ and $\lambda_0$ has been recently established experimentally in a series of Ba(Fe$_x$Ni$_{1-x}$)$_2$As$_2$ crystals  \cite{Rodiere}. A factor $\sim 3$ variation in $\Delta\lambda (5K)$ (for samples with similar $T_c$ values) should hence be directly related to a similar variation in $\lambda(0)$. However, the Ames group introduced the {\it same} $\lambda(0)$ value (measured in \cite{Kim}) to obtain $\rho_s(T)$ in \cite{Cho} and \cite{Kim} even though they measured a $\Delta\lambda (5K)$ value $\sim 3$ times smaller in \cite{Cho} than in \cite{Kim}. This inconsistent choice led by construction to very different temperature dependences of the superfluid density (see Fig.1) and forced the authors to drastically change their conclusion from "$\rho_s(T )$ at temperatures of the order of $T_c$ is fully described by only one component, determined by the band with a smaller gap" in \cite{Kim} to  "this result indicates that 75 \% contribution of superfluid density comes from the band which has the larger gap" in \cite{Cho}  shedding  doubts on the validity of any of those two conclusions  (see also discussion below).  

Note that the $\lambda(0)$ value used in \cite{Cho,Kim} is actually consistent with those obtained by other groups (using different techniques) : $\lambda_0 \sim 500$ nm $\pm 15$\% \cite{Klein,Biswas} and all samples (close to optimal doping) presented very similar $T_c$ values. As $\lambda(0)$ and $T_c$ \cite{Onari} are expected to be very sensitive to scattering, this suggests a rather good homogeneity in the different Fe(Se,Te) crystals. Similarly, a large number of specific heat measurements \cite{Cp,Klein} also led to very similar values of the jump at $T_c$ : $\Delta C_p/T_c \sim$ 40 mJ/molK$^2$ $\pm20$\% again confirming this homogeneity. It is hence difficult to attribute the dispersion in the $\Delta\lambda$ values to the sample bulk quality. However, as TDO measurements are sensitive to surfaces (for $T_c \rightarrow 0$), the influence of edge roughness - or any other surface inhomogeneity - can not be excluded. Note also that the slope of $\rho_s(t)$ for $t\rightarrow 1$ (with $t=T/T_c$) is thermodynamically related to $\Delta C_p$ and to the slope of the upper critical field ($B'_{c2}$) through $\Delta C_p=(\mu_0T_c).(dH_c/dT)^2_{|T\rightarrow T_c}\sim (B'_{c2}/\mu_0).(\Phi_0/4\pi\lambda_0^2).(d\rho_s/dt)_{|t\rightarrow 1}$. Introducing $B'_{c2} \sim 13$ T/K \cite{Klein}, one obtains $(d\rho_s/dt)_{|t\rightarrow 1}\sim 2.6$ ($\pm 40$\%) (see shaded area in Fig.1). As shown in Fig.1, the $\rho_s(T)$ data obtained in \cite{Cho} are in good agreement with those deduced from $H_{c1}$ measurements \cite{Klein} and $\mu$SR data \cite{Biswas} and thermodynamically consistent with the specific heat data whereas those previously obtained in \cite{Kim} (and TDO data in \cite{Klein}) clearly deviate from the thermodynamical cone.
 
\paragraph{On the fitting procedure.}  It is hence in principle possible to perform an analysis of the temperature dependence of the $\rho_s$ values obtained in \cite{Cho} (or more generally speaking on all the open symbols in Fig.1). However, the $T^n$ dependence measured at low temperature by all the groups strongly supports the presence of pair breaking effects. It is hence physically incorrect to fit the full temperature range dependence of this superfluid density using a two gap model in the {\it clean} limit.  Indeed, in this case, $\Delta\lambda$ should vary exponentially for  $k_BT < \Delta_{min}/5$, (where $\Delta_{min}$ is the smallest gap) that is for $T\leq3$ K taking the $\Delta_{min}$ value obtained in \cite{Cho}, in clear disagreement with their own data  displaying a $T^n$ law  in this temperature range. Note however that the influence of scattering on the temperature dependence of the superfluid density in a two gap system has been investigated in Ba(Fe,Co)$_2$As$_2$ by the Ames group in \cite{Kim2} and, since the temperature dependence observed in this latter system is very close to the one observed in Fe(Se,Te) it would have been much more valid to perform a similar analysis in \cite{Cho}. 

Finally, note that, $1/\lambda^2(T)$ is expected to vary as $1-t^2$  in the pair breaking model close to critical scattering (closed gap \cite{Gordon}). As shown in Fig.1, this dependence (solid line) reproduces very well the experimental dependence without any adjustable parameter whereas  {\it 6 parameters} (including $T_c$) have been used in \cite{Cho} to "self-consistently" fit the data. 

In conclusion, the origin of the large dispersion in the amplitude of $\Delta\lambda(T)$ obtained in Fe(Se,Te) crystals by TDO is still an open question but the possibility that edge roughness might alter the amplitude of $\Delta\lambda(5K)$ \cite{Hashimoto} can not be ruled out from the measurements presented in \cite{Cho}. Using the same $\lambda_0$ values to reconstruct the superfluid density in samples presenting very different $\Delta\lambda(T)$ (as done in \cite{Cho} and \cite{Kim}) is incorrect and leads by construction, to very different temperature dependences for $\rho_s$. Finally,  the power law dependence of the penetration depth at  low temperature ($\Delta\lambda \propto T^n$) is not compatible with the {\it clean} limit two-gap model used to describe the temperature dependence of the superfluid density in \cite{Cho}. The presence of strong scattering here hinderers any determination of the gap values from the temperature dependence of the superfluid density.

\end{document}